\documentstyle[12pt,a4wide]{article}
\input epsf

\newcommand{\be}{\begin{equation}}
\newcommand{\ee}{\end{equation}}
\newcommand{\bea}{\begin{eqnarray}}
\newcommand{\eea}{\end{eqnarray}}
\newcommand{\bean}{\begin{eqnarray*}}

\newcommand{\eean}{\end{eqnarray*}}
\font\upright=cmu10 scaled\magstep1
\font\sans=cmss12
\newcommand{\ssf}{\sans}
\newcommand{\stroke}{\vrule height8pt width0.4pt depth-0.1pt}
\newcommand{\Z}{\hbox{\upright\rlap{\ssf Z}\kern 2.7pt {\ssf Z}}}

\newcommand{\C}{{\rlap{\rlap{C}\kern 3.8pt\stroke}\phantom{C}}}
\newcommand{\R}{\hbox{\upright\rlap{I}\kern 1.7pt R}}
\newcommand{\CP}{\C{\upright\rlap{I}\kern 1.5pt P}}
\newcommand{\PP}{\hbox{\upright\rlap{I}\kern 1.5pt P}}

\newcommand{\identity}{{\upright\rlap{1}\kern 2.0pt 1}}

\newcommand{\pibf}{\mbox{\boldmath $\pi$}}
\newcommand{\taubf}{\mbox{\boldmath $\tau$}}

\begin{document}
\pagestyle{plain}
\title{\vskip -70pt
\begin{flushright}
{\normalsize UKC/IMS/97-07} \\
{\normalsize IMPERIAL/TP/96-97/18} \\
\end{flushright}\vskip 50pt
{\bf \Large \bf SYMMETRIC SKYRMIONS}\vskip 10pt}
\author{Richard A. Battye$^{\ \dagger}$ and Paul M. Sutcliffe$^{\ \ddagger}$\\[10pt]
{\normalsize $\dagger$ {\sl Theoretical Physics Group, Blackett Laboratory, Imperial College,}}\\{\normalsize {\sl Prince Consort Road, London, SW7 2BZ, UK.}}\\{\normalsize {\sl Email : R.Battye@ic.ac.uk}}\\ 
\\{\normalsize $\ddagger$ {\sl Institute of Mathematics, University of Kent at Canterbury,}}\\
{\normalsize {\sl Canterbury, CT2 7NZ, U.K.}}\\
{\normalsize{\sl Email : P.M.Sutcliffe@ukc.ac.uk}}\\}
\date{}
\maketitle
\vskip 25pt
\begin{abstract}
\noindent We present candidates for the global minimum energy solitons of 
charge one to nine in the Skyrme model, generated using sophisticated numerical
 algorithms. Assuming the Skyrme model accurately represents the low energy
 limit of QCD, these configurations correspond to the classical nuclear ground
 states of the light elements. 
The solitons found are particularly symmetric, for example, the charge seven
skyrmion has icosahedral symmetry, and 
the shapes are shown to fit a
 remarkable sequence defined by a geometric energy minimization (GEM) rule.
 We also calculate the energies and sizes to within at least a few percent accuracy.
 These calculations provide the basis for a future investigation of the low 
energy vibrational modes of skyrmions and hence the possibility of testing
 the Skyrme model against experiment.
\end{abstract}
\newpage

The Skyrme model presents an opportunity to understand nuclear physics as a low
 energy limit of quantum chromodynamics (QCD). It was first proposed as a theory
 of strong interactions of hadrons\cite{S}, but after a hiatus which lasted 
nearly twenty years it was shown to be in fact the low energy limit of QCD
 in the large $N_c$ limit\cite{WIT}. Since then much work has suggested that 
the topological solitons in the model, known as skyrmions, can indeed be 
identified with classical ground states of light nuclei. However, before a
 more quantitative assessment of the validity of the model can be made, a 
 thorough understanding of the structure and dynamics of multi-soliton 
configurations is required.

This is a highly non-trivial task since the Skyrme model has many features 
which obstruct both analytic and numerical progress. There are two promising
 lines of attack. The first is to use the approximate correspondence between
 Skyrme fields and SU(2) instantons\cite{AMa}. Using this 
 semi-analytic approach, one can understand the right-angle scattering of two skyrmions\cite{AMb}
and also calculate symmetric configurations of, for example, charge three  and four\cite{LM}.
 The second is to use state of the art computer technology and a sophisticated
 numerical algorithm to evolve the dynamics of discretized configurations. This approach has the
 advantage of being
quantitatively accurate, but has many other more practical difficulties associated with it,
 in particular the substantial computer resources required. Nonetheless, it has been possible to
 design a working code which can evolve the dynamics of low energy configurations for many timesteps
 and also be modified to relax dynamically towards what are at least local minimum energy 
configurations. 

In a recent letter\cite{BSuta}, we presented the first results
 from our code, exhibiting the dynamics 
of symmetric configurations in the attractive channel and also
 the minima for charges one to four. 
These minima were already known from numerical  relaxation 
calculations\cite{BTC} and also from 
instanton calculations\cite{LM}. In this letter, we present 
what we describe as candidate minima for
 charges one to nine, which we firmly believe are in fact the 
global minima. These minima can be
 classified in terms of the isosurfaces of baryon density (or
 energy density), which are seen to fit
 a remarkable pattern, obeying what we shall call a geometric
 energy minimization (GEM) rule. 
We also calculate accurately the energy and average size of
 each soliton. As in ref.\cite{BSuta},
 we shall assume the pion mass is zero --- results for a 
physical value of the pion mass will be
 presented in future work. We do not expect
 that inclusion of a finite pion mass will
 effect the overall shape and symmetry of the soliton,
 rather it will just modify its energy and size.
  
The Lagrangian of the massless Skyrme model may be written in terms
of the SU(2) valued right currents $R_\mu=(\partial_\mu U)U^\dagger$ as
$$
12\pi^2{\cal L} =-\textstyle{1\over 2}
\mbox{Tr}(R_\mu R^\mu)-\textstyle{1\over 16}\mbox{Tr}
([R_\mu,R_\nu][R^\mu,R^\nu])\,
\label{lag}
$$
where $U=\sigma+i{\taubf}\cdot{\pibf}$ and we have used scaled units of energy, 
$E=661{\rm MeV}$, 
and length, $\ell=0.755{\rm fm}$. One can convert to an O(4) sigma model
 representation
 $\phi=(\sigma,\pi_1,\pi_2,\pi_3)$ as described in refs.\cite{BSuta,BSutb} and
 deduce the dynamical
 equations of motion for $\phi$. The baryon density ${\cal B}$, whose spatial
 integral gives the
 integer-valued baryon number $B$, is given by
$24\pi^2{\cal B}=-\epsilon_{ijk}\mbox{Tr}(R_iR_jR_k)$
where Latin indices  run over the
spatial values $1,2,3$.
The above units are chosen so that the Fadeev-Bogomolny bound on the 
energy $E$ is simply $E\geq |B|$.

The full details of the numerical methods are presented in ref.\cite{BSutb}. 
This includes an account of how the numerical algorithm works, how to construct
 the initial conditions and why
 we believe that the configurations represent the global minima. For the purposes of this letter,
 we include a brief outline for the layman. The basic procedure is to set up discretized initial 
conditions for static low energy configurations using either a simple ansatz for a single skyrmion
 and the product ansatz, or by calculating the holonomy of instantons, or a combination of both, 
as is the case for the higher charge configurations. The critical feature of the specific
 configurations chosen is that the skyrmions are in an attractive or nearly attractive channel.
 These configurations are then evolved using the discretized equations of motion. As the dynamics
 proceeds the system oscillates between maxima and minima of the potential energy. If one stops the
 dynamics at a minimum of the potential energy and then removes all the kinetic energy from the system, that is, set the time derivatives of the field to zero, the solution will gradually move 
towards a local minimum of the system. Once the system is close to the minimum one can also 
incorporate dissipation which will speed up the process of relaxation. Of course, as with any numerical
 minimization procedure, one cannot be sure that one has found the global minimum. However,
 given sufficiently asymmetric, but attractive, initial conditions, as created using the product 
ansatz, it is likely that one will locate the global minimum.

For charges one to four it was possible to construct configurations using just the product ansatz,
 in which all the skyrmions are in a mutually attractive channel\cite{BSuta,BSutb}. It is not
 possible to use this naive approach to construct a maximally attractive channel of charge five 
or higher and therefore one must consider other approaches. In order for our algorithm to relax 
quickly to the minimum, we must construct a configuration with low energy, in which most of the 
skyrmions are attractive. It need not be the maximally attractive channel, but the algorithm will
 relax to the minimum quicker if it is close to the most attractive. Some 
of the
 configurations chosen relax quicker than others, suggesting that in some cases we have selected the correct
 configuration and in others we have found one which works, but is perhaps not the best.

We find that it is best to mix the instanton approach for a known symmetric 
configuration, with a small number of single skyrmions added using the product 
ansatz to break the exact symmetries. For example, in the case of $B$=$5$, it
 possible to construct a highly attractive configuration by adding in two single
 skyrmions either side of a $B$=$3$ tetrahedron. Since the pion fields of the 
tetrahedron are similar to those of an anti-skyrmion at large distances, the 
configuration is of low energy and relaxes extremely quickly ($\sim$1000 timesteps)
 to the minimum. Using the analogy, to $B$=$2$ to $B$=$4$ scattering, we tried
 constructing $B$=$6$ and $B$=$7$ configurations by surrounding a $B$=$3$ tetrahedron 
with skyrmions in cyclic and tetrahedral configurations respectively using the 
product ansatz. These configurations relaxed toward the minimum in over 5000 
timesteps, suggesting that in fact these are not the maximally attractive channel.
 Nonetheless, they are of reasonably low energy and work eventually.
 For $B$=$6$, one can relax to the minimum much more quickly by colliding two 
tetrahedra. Reassuringly, it is the same minima as calculated before, providing 
a useful consistency check. In order to calculate configurations of higher charge 
$(B>7)$, one can just add in a single skyrmion to the known minima of charge one less.

A useful way to represent a skyrmion is by displaying a surface of constant baryon density. In fig.~1 we display isosurfaces
for the skyrmions of charge five to nine (one to four are presented in ref.\cite{BSuta}), using the same constant value for the baryon density
in each case, to ensure that the relative sizes of the skyrmions are accurately represented.
The baryon density has maxima at several points in space, which we can think of as vertices
of a solid, and these are connected by links of slightly lower baryon density, which can
be regarded as the edges of the solid. Clearly in this way we can assign a solid to each skyrmion
which accurately reflects the shape and symmetry of the skyrmion. In fact, for all the skyrmions
we consider, the associated solid is remarkably close to being composed of regular or nearly regular polygons with
a fixed edge length. Given this fascinating result, we shall describe each solid in terms
of its construction from polygons. Included in fig.~1 alongside each baryon density plot
is a photograph of the associated solid, constructed using a molecular model building kit.
In table 1 we list the number of faces of each associated solid, together with the number of
each type of polygon from which it is constructed.

\medskip
\begin{center}
\begin{tabular}{|c|c|c|c|c|c|c|} \hline
B& Faces & Triangles & Squares & Pentagons & Hexagons & Symmetry \\
\hline
3 &  4 & 4 & 0 & 0  & 0 & $T_{d}$ \\
4 &  6 & 0 & 6 & 0  & 0 & $O_{h}$ \\
5 &  8 & 0 & 4 & 4  & 0 & $D_{2d}$ \\
6 & 10 & 0 & 2 & 8  & 0 & $D_{4d}$ \\
7 & 12 & 0 & 0 & 12 & 0 & $I_{h}$ \\
8 & 14 & 0 & 0 & 12 & 2 & $D_{6d}$ \\
9 & 16 & 0 & 0 & 12 & 4 & $T_{d}$ \\
\hline
\end{tabular}
\end{center}
{\bf Table 1 } : The constituent polygons and symmetry structure for the candidate minima of charge three and above.
\medskip

It is perhaps useful to give a brief description of each of the solids, which in conjunction
with fig.~1,  should make the structure of each configuration clear. The $B$=$5$ skyrmion consists
of two parallel down-pointing pentagons attached to two more parallel up-pointing pentagons, so
that four sides of a box are formed. The top of the box is formed by adding two squares, and
similarly for the bottom of the box, though of course the arrangement of joined squares
on the top and bottom have a relative rotation of $90^\circ$. The $B$=$6$ configuration consists of
 two halves, each
of which is formed from a square with a pentagon hanging down from each of its four sides.
Note that to join these two halves implies that the two squares are parallel, but one is rotated
by $45^\circ$ relative to the other. The $B$=$7$ solid is a regular dodecahedron. 
The $B$=$8$ skyrmion has a similar structure to its $B$=$6$ counterpart, except that the squares
are replaced by hexagons, so that each half has six pentagons hanging down. This requires
the top hexagon to be parallel to the bottom hexagon but rotated by $30^\circ$.
Finally, the $B$=$9$ structure has four hexagons located at the vertices of a regular tetrahedron
which are joined by four sets of three connected pentagons, whose single common vertex
lies at the vertices of the dual tetrahedron.

The symmetries of each skyrmion are of interest and here we shall discuss the symmetry group
of the baryon density (or equivalently energy density). To discuss the symmetries of the field
itself is a more complicated task, since the three-dimensional representation of the symmetry
group which acts on the pion fields as an isospin rotation must also be identified.
In the final column of table 1 we give the symmetry group, where we use the 
Sch\"onflies notation (see for example, ref.\cite{Sch}) popular in chemistry.
All the configurations contain at least the symmetry group $D_{nd}$, for some $n$.
This symmetry group is obtained from the cyclic group of order $n$, $C_n$, by the addition
of a $C_2$ symmetry with axis perpendicular to the main symmetry axis and a reflection
symmetry in a vertical plane containing the main axis and which bisects the angle
 between
pairs of $C_2$ axes. Thus these twisted dihedral symmetries (which are enhanced to platonic
symmetry in some cases) appear to be of importance to skyrmions, as
they are for BPS monopoles\cite{HSb,HSc}.

Note that the skyrmions of charge 3,4,7 have the same platonic forms as the
 corresponding
BPS monopoles of the same charge\cite{HMM,HSa,HSb}, but that the charge 5
 configuration
is not an octahedron, even though an octahedral 5-monopole exists\cite{HSb}.
 An octahedral
charge 5 skyrmion does exists but it is not the minimal energy 
configuration\cite{BSutb}. This illustrates an important difference
 between BPS monopoles and skyrmions. All the monopole configurations of a 
particular charge have the same energy and 
it is due to a mathematical simplification that
 only the very symmetric ones
 have been found, but the skyrmions evolve under the influence of a potential 
allowing for the possibility of minima. Nonetheless, it is fascinating that in
 reality the highly symmetric low energy configurations of $B$=$3$, $B$=$4$, 
$B$=$7$ and $B$=$9$
form in the early universe during Big Bang Nucleosynthesis,
 while the less symmetric  configurations of $B$=$5$, $B$=$6$ and $B$=$8$ 
are either unstable or barely stable.

Given the complex structure of these skyrmions the question now arises as whether a rule
exists which fits the remarkable sequence of shapes found. We propose the following
phenomenological rule for the structure of the  minimum energy charge $B>2$ skyrmion, 
which we refer to as the Geometric Energy Minimization (GEM) rule;\\

\noindent {\sl GEM RULE} : 
{\sl The charge $B$ baryon density surface is composed of almost regular polygons and consists of
 $4(B-2)$ trivalent vertices. If more than one such solid exists, then select the most spherical.}\\

It should be noted that there are several equivalent ways in which the GEM rule could be stated. For example, from the trivalent property together with Eulers formula, fixing one of the three parameters of the solid, that is, the number of vertices,
 faces and edges, determines the other two. Explicitly we have
$V=4(B-2),\ F=2(B-1), E=6(B-2)$. Thus since the baryon density isosurface has a hole in the centre of
each face then the GEM rule implies the observation of ref.\cite{BTC} that the isosurface contains
$2(B-1)$ holes. As the value of $B$ increases the number of possible configurations satisfying
the first part of the GEM rule grows, and hence the need for the second part. For example, at
$B=6$ in addition to the configuration found there is a second possibility that consists
of a hexagon with alternating squares and pentagons rising from its edges, and topped by
three joined pentagons. This crown-like configuration, which has cyclic $C_3$
 symmetry,
 satisfies the first part of the GEM rule,
but is not very spherical, having a flat bottom and a pointed top. It may be that some other
statement, such as a minimization of the standard deviation of edge lengths, or a similar
property for edge angles, is an improved statement
of the spherical property. More examples of higher charge skyrmions are required to resolve
this issue.

It follows from the GEM rule that for $B\ge 7$ the solid consists of $12$ 
pentagons
and $2(B-7)$ hexagons. Such configurations occur in fullerene chemistry\cite{Bag,Kro}, where
we should compare with the carbon structure $C_{4(B-2)}$. In fullerine chemistry avoiding
large curvature is important, so the first fullerene is $C_{28}$, avoiding
a fused quartet of pentagons, which has precisely the form of the $B$=$9$
skyrmion. It would be interesting to see if this correspondence continues, 
since the $B$=$17$ skyrmion should have the $C_{60}$ Buckminsterfullerene structure.
 Extension of our current results will require considerable amounts of computer resources,
 but this is now in progress. 

\medskip
\begin{center}
\begin{tabular}{|c|c|c|c|c|c|c|c|c|c|} \hline
B&  $B_{\rm dis}$ & $E_{\rm dis}$ & $E_{\rm dis}/B_{\rm dis}$ & $E$ & $I$ & $\Delta r$ &  $E$(MeV) & $I$(MeV) & $\Delta r$(fm)\\
\hline
1 & 0.984 & 1.212 & 1.232 & 1.232 &       & 1.034 &  814 &     & 0.781 \\
2 & 1.972 & 2.308 & 1.171 & 2.342 & 0.122 & 1.416 & 1547 &  81 & 1.092 \\
3 & 2.960 & 3.384 & 1.143 & 3.429 & 0.145 & 1.636 & 2267 &  94 & 1.235 \\
4 & 3.948 & 4.407 & 1.116 & 4.464 & 0.197 & 1.860 & 2951 & 130 & 1.404 \\
5 & 4.935 & 5.509 & 1.116 & 5.580 & 0.116 & 2.035 & 3689 &  76 & 1.536 \\
6 & 5.923 & 6.567 & 1.109 & 6.654 & 0.158 & 2.220 & 4396 & 107 & 1.676 \\
7 & 6.913 & 7.596 & 1.099 & 7.693 & 0.193 & 2.332 & 5083 & 127 & 1.761 \\
8 & 7.900 & 8.690 & 1.100 & 8.800 & 0.125 & 2.487 & 5816 &  81 & 1.878 \\
9 & 8.885 & 9.891 & 1.099 & 9.891 & 0.141 & 2.623 & 6534 &  96 & 1.980 \\
\hline
\end{tabular}
\end{center}
{\bf Table 2 } : Calculated values of the charge ($B_{\rm dis}$),
 energy ($E_{\rm dis}$) and soliton size ($\Delta r$) in natural
 and physical units for $B=1$ to $B=9$. Also presented is the 
 ionization energy $I$, which is the energy required to remove one skyrmion.
 For a discussion of the accuracy of the results and comparison to previous
 calculations see the text.
\medskip

We have calculated the discrete charge and energy for the minima on $100^3$ grids,
 which are displayed in table 2. These energy values are less than the 
true values since the grid size is finite, but they are nonetheless within
 2\% accurate. One can make a better estimate of the overall energy 
by using the ratio $E_{\rm dis}/B_{\rm dis}$ as discussed in ref.\cite{BSuta}. This can be seen to be exact
 to 3 decimal places for the $B$=$1$ skyrmion and we suggest that it will be so 
for all the others\footnote{It should be noted that the quoted values for $B$=$1$
 to $B$=$4$ differ very slightly from those presented in ref.\cite{BSuta}.
 The current values are the result of further relaxation of the same
 configurations.}, since the third decimal place has not changed for many 
($>$1000) timesteps. These values have been used to calculate the ionization 
energy ($I_B=E_1+E_{B-1}-E_B$), that is, the energy required to remove one
 skyrmion, which in general gives an indication of the classical binding energy.
 Taking into account the fact that these energies will be modified by the
 quantisation of zero modes and also some low energy vibration modes, there is
 a remarkable trend which follows the pattern of light element binding energies.
We have also calculated the average size of the soliton $\Delta r$ from the second moment of the baryon distribution as in ref.\cite{BTC}. This value is extremely rough since it ignores the symmetries of the object, but nonetheless it gives an indication of the overall trend.

We should comment on the relevance of our work to previous calculations. In numerical work\cite{BTC}, similar to ours, configurations up to charge six were
studied using a global minimization algorithm. Our results for charges five and six differ from these earlier computations,
which we attribute to numerical effects in the earlier work due  to lack of resources; for example, the grid we use
contains almost five-times the number of points.
For charge five the difference is small, we obtain the same symmetry group but
identify different polygons forming the solid, which is possible thanks to our improved
grid resolution. For charge six our results are very different, as
we find a very symmetric configuration, whereas the earlier computation gave a structure
with very little symmetry.

A different approach to constructing high charge configurations 
has used the Skyrme crystal\cite{KS} and involves cutting out
sections\cite{kim}. Although these configurations have low energy,
 they are not as low as those presented here, and they are fundamentally 
different in nature. The configurations presented here are shells with
 less baryon density in the centre, whereas those created from the Skyrme crystal
 have internal structure since they are created from cubic configurations.
 Such structures do not fit the GEM rules since they are not trivalent.
It is an open question as to whether the shell structure persists for higher charge.

We believe that the candidate minima which we have presented here are in fact 
the global minima since the initial conditions have natural asymmetry, we have 
in some cases got the same configuration from two different initial conditions and
 most of all the isosurfaces of the baryon density fit a remarkable sequence of
 polygons. The symmetry properties of these polygons could be the starting point
 for an understanding of the moduli space structure of the Skyrme model, 
leading to a study of low energy dynamics. The physical properties of these
 classical nuclear ground states  follow the trend of light elements, at least 
qualitatively, providing the impetus for a study of the low energy vibrational
 modes and the potential for a confrontation between the Skyrme model and nuclear
 experiments.\\

\noindent We have benefited from useful conversations with Nick Manton,
 Conor Houghton, Brad Baxter, 
Paul Shellard, Jonathan Moore, Dick Hughes-Jones, Neil Turok and Kim Baskerville. 
RAB acknowledges the support of PPARC Postdoctoral fellowship grant GR/K94799 
and PMS thanks the Nuffield Foundation for a newly appointed lecturer award. 
We also acknowledge the use of the SGI Power Challenge at DAMTP in Cambridge 
supported by the PPARC Cambridge Relativity rolling grant and EPSRC Applied 
Mathematics Initiative grant GR/K50641. Thanks to Paul Shellard for his 
understanding in these respects.

\def\jnl#1#2#3#4#5#6{\hang{#1 [#2], {\it #4\/} {\bf #5}, #6.}}
\def\jnltwo#1#2#3#4#5#6#7#8{\hang{#1 [#2], {\it #4\/} {\bf #5}, #6;{\bf #7} #8.}}
\def\prep#1#2#3#4{\hang{#1 [#2],`#3', #4.}} 
\def\proc#1#2#3#4#5#6{{#1 [#2], in {\it #4\/}, #5, eds.\ (#6).}}
\def\book#1#2#3#4{\hang{#1 [#2], {\it #3\/} (#4).}}
\def\jnlerr#1#2#3#4#5#6#7#8{\hang{#1 [#2], {\it #4\/} {\bf #5}, #6.
{Erratum:} {\it #4\/} {\bf #7}, #8.}}
\def\prl{Phys.\ Rev.\ Lett.}
\def\pr{Phys.\ Rev.}
\def\pl{Phys.\ Lett.}
\def\np{Nucl.\ Phys.}
\def\prp{Phys.\ Rep.}
\def\rmp{Rev.\ Mod.\ Phys.}
\def\cmp{Comm.\ Math.\ Phys.}
\def\mpl{Mod.\ Phys.\ Lett.}
\def\apj{Ap.\ J.}
\def\apjl{Ap.\ J.\ Lett.}
\def\aap{Astron.\ Ap.}
\def\cqg{Class.\ Quant.\ Grav.} 
\def\grg{Gen.\ Rel.\ Grav.}
\def\mn{M.$\,$N.$\,$R.$\,$A.$\,$S.}
\def\ptp{Prog.\ Theor.\ Phys.}
\def\jetp{Sov.\ Phys.\ JETP}
\def\jetpl{JETP Lett.}
\def\jmp{J.\ Math.\ Phys.}
\def\zpc{Z.\ Phys.\ C}
\def\cupress{Cambridge University Press}
\def\pup{Princeton University Press}
\def\wss{World Scientific, Singapore}
\def\oup{Oxford University Press}

\section*{Figure Captions}  

\noindent Fig.~1. skyrmions of charge five to nine; on the left baryon density 
isosurfaces (to scale) with five at the top and nine at the bottom and on
 the right wireframe models of the corresponding solids (not to scale).

\bigskip

\end{document}